\documentclass[useAMS,usenatbib]{mn2e}

\usepackage{amsmath}
\usepackage{graphicx}
\usepackage{ulem} 

\newcommand{\eqb}{\begin{equation}}  
\newcommand{\eqe}{\end{equation}}			
\newcommand{\bl}{B_{\rm L}}
\newcommand{\rl}{r_{\rm L}}
\newcommand{\gammav}{\Gamma}
\newcommand{\betav}{\beta}

\newcommand{\apj}{ApJ}
\newcommand{\apjs}{ApJS}
\newcommand{\apjl}{ApJL}
\newcommand{\mnras}{MNRAS}
\newcommand{\aap}{A\&A}
\newcommand{\nat}{Nature}

\title[VHE photons from reconnection in pulsar winds]{Very high energy emission as a probe of relativistic 
magnetic reconnection in pulsar winds}
\author[I. Mochol and J. P\'etri]{Iwona Mochol$^{1,2}$ and J\'er\^{o}me P\'etri$^1$ \thanks{E-mail: iwona.mochol@astro.unistra.fr; jerome.petri@astro.unistra.fr} 
\\
$^{1}$ Observatoire Astronomique de Strasbourg, Universit\'e de Strasbourg, CNRS, UMR 7550, 11 rue de l'Universit\'e, F-67000 Strasbourg, France. \\
$^{2}$ Institute of Nuclear Physics, Polish Academy of Sciences, ul. E. Radzikowskiego 152, 31-342 Krak\'ow, Poland}
\begin{document}

\date{Accepted . Received ; in original }

\pagerange{\pageref{firstpage}--\pageref{lastpage}} \pubyear{2014}

\maketitle

\label{firstpage}

\begin{abstract}
The population of gamma-ray pulsars, including Crab observed in the TeV range, and Vela detected above 50~GeV, challenges existing models of pulsed high-energy emission. Such models  should be universally applicable, yet they should account for spectral differences among the pulsars.  

We show that the gamma-ray emission of Crab and Vela can be explained by synchrotron radiation from the current sheet of a striped wind, expanding with a modest Lorentz factor $\Gamma\la100$ in the Crab case, and $\Gamma\la50$ in the Vela case. In the Crab spectrum a new synchrotron self-Compton component 
is expected to be detected by the upcoming experiment CTA. 

We suggest that the gamma-ray spectrum directly probes the physics of relativistic magnetic reconnection in the striped wind. In the most energetic pulsars, like Crab, with $\dot{E}_{38}^{3/2}/P_{-2}\ga0.002$ (where $\dot{E}$ is the spin down power, $P$ is the pulsar period, and $X=X_i\times10^i$ in CGS units), reconnection proceeds in the radiative cooling regime and results in a soft power-law distribution of cooling particles; in less powerful pulsars, like Vela,  particle energization is limited by the current sheet size, and a hard particle spectrum reflects the acceleration mechanism. 
{A strict lower limit on the number density of radiating particles corresponds to emission close to the light cylinder, and, in units of the GJ density, it is $\ga0.5$ in the Crab wind, and $\ga0.05$ in the Vela wind.}
\end{abstract}

\section{Introduction}

It is now well established that the pulsed spectrum of the Crab extends up to 400 GeV \citep{2011Sci...334...69V,2012A&A...540A..69A} and very recently new measurements of TeV photons have been announced \citep{crabtev}. Although this is the most extreme example, several other pulsars exhibit emission above 25 GeV \citep{2013ApJS..209...34A},   
including the Vela pulsar with reported detection above 50 GeV \citep{2014arXiv1410.5208L}.  
The Fermi-LAT data of these objects have been fitted with a sub-exponential cut-off or a broken power-law. 

Such emission has been modelled as inverse Compton scattering of magnetospheric photons by the $e^{\pm}$ pairs created and accelerated in the magnetosphere \citep{2013MNRAS.431.2580L}. As an alternative, wind models have been proposed, which attribute the gamma-ray emission to the synchrotron radiation of energized plasma either very close to the light cylinder \citep{1996A&A...311..172L, 2010ApJ...715.1282B, 2013A&A...550A.101A}, or in the far wind zone \citep{2012MNRAS.424.2023P}. Inverse Compton scattering of the thermal stellar photons by the instantaneously accelerated wind has been also considered in this context \citep{2012Natur.482..507A}.

In the standard wind scenario (the \lq\lq striped wind\rq\rq\  model), the bulk speed of the flow is roughly constant and highly relativistic.  The entire physics of pulsed emission is related to strong beaming effects and the presence of current sheets \citep{2002A&A...388L..29K} -- narrow regions in the wind, which are embedded in the cold, highly magnetized stripes of opposite magnetic polarity \citep{1990ApJ...349..538C}. 
 Radiating particles are thought to be energized during reconnection in the current layer and they are thought to have relativistic thermal distribution that satisfies a Harris equilibrium \citep{2005ApJ...627L..37P}. However, recent simulations of relativistic reconnection in plasmas with high magnetization, $\sigma\gg1$ (where $\sigma$ is the ratio of Poynting flux to bulk kinetic-energy flux), suggest that the energized particles are accelerated to a power-law distribution 
\citep{2014ApJ...783L..21S, 2014arXiv1409.8262W}.  
Moreover, it has been argued that in highly magnetized, relativistic systems shearing of magnetic field lines leads to formation of nearly force-free current layers rather than hot plasma sheets,  
and therefore a force-free equilibrium is a more appropriate model than a Harris equilibrium \citep[and references therein]{2014PhRvL.113o5005G}. 

In pulsar electrodynamics, the formation of a current layer can start close to the neutron star surface, along the last open field lines, and extend far outside the light-cylinder into the striped wind region. 
We describe a current layer as approximately force-free up to the light cylinder, beyond which reconnection sets in and proceeds over many wavelengths in the wind as it propagates. Since reconnection results in energy dissipation and acceleration of particles, the energy density in the particle content increases and, together with the reconnecting field in a sheet, it maintains the pressure balance against the fields outside the layer. 
This approach can be compared with the FIDO model (force-free inside -- dissipative outside the light cylinder) of \citet{2014ApJ...793...97K}, who, however, attribute the energy dissipation to a changing plasma conductivity. They have demonstrated that the model is able to reproduce the variety of Fermi-LAT pulsar lightcurves.

We focus on the spectra of gamma-ray pulsars.  
In our model the spectral shape is an imprint of the acceleration regime, which can be    
determined either by the radiative cooling of particles or by the size of the acceleration region (current sheet thickness). 
We show that the wind scenario can be distinguished from other competing models by observations of its unique signature: synchrotron self-Compton (SSC) emission extending up to tens of TeV in the spectrum of the Crab pulsar. 

In section~\ref{sec:Model}, we describe the basic picture of our wind model, including the magnetic configuration and the particle distribution function. Next, as typical examples of two acceleration regimes we investigate the emission properties of the winds from the Crab pulsar and the Vela pulsar, as explained in section~\ref{sec:Emission}. Conclusions are drawn in section \ref{sec:Conclusion}.

\section{The model}
\label{sec:Model}
\subsection{Magnetic field structure}

The wind description is based on the striped wind model of \citet{2013MNRAS.434.2636P}, 
valid beyond the light cylinder. In this model the wind (fluid) expands with a {constant} relativistic Lorentz factor $\gammav=(1-\betav^2)^{-1/2}$ in the radial direction. 
In the wind comoving frame the electric field vanishes. We formulate the relevant physics 
in this frame and all the quantities in this frame are denoted with a prime, while non primed quantities should be understood as expressed in the lab frame. 

In the fluid comoving frame the magnitude of the magnetic field is
\begin{align}
B'^2&=\betav^2\bl^2\frac{\rl^2}{r^2}\left(\frac{\sin^2\theta}{\gammav^2}+\betav^2\frac{\rl^2}{r^2}\right)\tanh^2(\psi_{\rm s}/\Delta) \nonumber \\
&=B'^2_{\rm max}\tanh^2(\psi_{\rm s}/\Delta)
\label{stripecomp}
\end{align}
where $\psi_{\rm s}=\cos\theta\cos\chi + \sin\theta\sin\chi\cos\left[\phi-\omega(t-r/\betav c)\right]$ describes the location of the current sheet of the wind, with $\chi$ being the angle between magnetic and rotational axes (obliquity), $\omega=2\pi/P$ being the pulsar frequency, $P$ -- the pulsar period, and $\rl=c/\omega$ the light-cylinder radius. {The current sheet thickness is parametrised by $\lambda \, \Delta$, where $\lambda= 2\pi \beta\,r_{\rm L}$ is the striped wind wavelength and $\Delta\in[0;0.5]$.} 

The ordered field exists mostly in stripes, vanishing in the middle of a current sheet.
We assume, however, that the current sheet is magnetized such that there is an additional field component, generated in the course of reconnection, which can be either shearing or random, and which we call \lq\lq turbulent\rq\rq. {There are two reasons for invoking this component: (1) a nonvanishing synchrotron emissivity in the middle of the current sheet (where the ordered component vanishes), (2) the turbulent nature of the field seems to be required to explain the decrease in linear polarization degree of the Crab optical pulse} \citep{2013MNRAS.434.2636P}. 

We assume that only a fraction $\varepsilon_{\rm d}$ of the available magnetic energy is dissipated during reconnection  
and most of the field survives as a turbulent field.  
The simplest model for this turbulent component proves to be 
$B'_{\rm tur}=b_0 B'_{\rm max}/\cosh(\psi_{\rm s}/\Delta)$,  
{where $b_0=(1-\varepsilon_{\rm d})^{1/2}$.}
This choice ensures that in the case of no dissipation ($b_0=1$) a turbulent field provides a pressure balance between a current sheet and the field outside. 
The energy dissipation and particle energization may be interpreted as a change in the fraction of plasma and field contributions to the pressure balance \citep{2009PhRvL.102m5003H}, which, in the wind comoving frame, takes the form
\eqb
\frac{B'^2_{\rm max}}{8\pi}=\frac{B'^2}{8\pi}+p+\frac{B'^2_{\rm tur}}{8\pi} \label{pbalance}
\eqe  
The pressure of ultrarelativistic particles can be approximated by $p\approx u'/3$, where $u'=mc^2 \int \gamma n'(\gamma) d\gamma$ is the energy density and $n'(\gamma)$ is the distribution function. 
\subsection{Particle distribution function}
 
So far the current sheet in a striped wind has been modelled by assuming a thermal particle distribution with a temperature determined by the synchrotron cooling  \citep{2012MNRAS.424.2023P}. However, simulations suggest that during reconnection particles are energized to a broader power-law distribution. 
Power law indices $s\approx1$ are attributed to the acceleration at the primary X-point, whereas indices up to $s\approx3$ result from acceleration in many X-points \citep{2004PhPl...11.1151J}. These spectral features depend also on the plasma magnetization $\sigma$ \citep{2014PhRvL.113o5005G, 2014arXiv1409.8262W}, such that for the highest values of magnetization the spectral index tends to one, $s\rightarrow1$. Radiative cooling has not been probed in simulations, but we argue that in this case a cooled distribution may be observed if the acceleration and cooling timescales are shorter than the timescale of particle injection into the acceleration process.  

We suggest that depending on the pulsar period $P$ and spin down power $\dot{E}$, 
particle acceleration at reconnection sites in the wind proceeds in two different regimes. 

In the first -- cooling regime -- the particle spectrum is modelled as a power law with an exponential cut off: 
\eqb
n'(\gamma)=n'_0\gamma^{-s}e^{-\gamma/\gamma'_{\rm rad}} 
\label{distrrad}
\eqe 
$\gamma'_{\rm rad}$ is the particle Lorentz factor, for which the synchrotron cooling timescale in the total magnetic field $B'_{\rm tot}=(B'^2+B_{\rm tur}'^2)^{1/2}$ is equal to the acceleration timescale in the reconnection electric field: 
\begin{align}
\gamma'_{\rm rad}&=\left(\frac{6\pi e\tau}{\sigma_{\rm T}B'_{\rm tot}}\right)^{1/2} \\
&\approx3.4\times10^5 \Gamma_{1}^{3/2}P_{-2}^{1/2}\varepsilon_{\rm d,-2}^{1/2}\dot{E}_{38}^{-1/4} \left(\frac{1+\sin^2\chi}{\Gamma_1^2\hat{r}^{-2}_1+\sin^2\theta}\right)^{1/4}
\label{gammarad}
\end{align}
where we use $\bl\approx(2\pi/P)(\dot{E}/c^3)^{1/2}(1+\sin^2\chi)^{-1/2}$ \citep{2006ApJ...648L..51S} and the notation $X=X_i\times10^i$ in CGS units. 
{We have also assumed} that the reconnection electric field has a form $E'\approx \tau B'$, where $\tau\la1$ is a constant reconnection rate. 
We estimate it by comparing the reconnection timescale $t'_{\rm rec}\sim\gammav \rl/(\tau c)$ with the wind expansion time $t'_{\rm exp}\sim r_{\rm diss}/(\varepsilon_{\rm d} \gammav c)$ \citep{2013MNRAS.431..355G}, where $\epsilon_{\rm d}\ll 1$ determines the fraction of the available magnetic energy that is dissipated within the distance $r_{\rm diss}$. Thus, the reconnection rate can be related to the dissipation distance $\tau\approx\gammav^2\varepsilon_{\rm d}\rl/r_{\rm diss}$ (the dissipation distance {in units of the light cylinder radius, $\hat{r}=r_{\rm diss}/\rl$,} is a parameter in our model, and we obtain it by fitting of the computed synchrotron flux to the pulsar spectrum, see Sect.~\ref{sec:fitting}). 
For usually invoked reconnection rates ($\tau\sim0.01-0.2$) the full dissipation ($\varepsilon_{\rm d}\approx1$) would proceed over many wavelengths in the wind, possibly accelerating it \citep{2001ApJ...547..437L}. The assumption $\varepsilon_{\rm d}\ll1$ allows us to neglect the wind acceleration. 

In the second regime, acceleration of particles is limited by the current sheet size, thus they escape the acceleration region before they reach the cooling regime. {This case has been investigated by} \citet{2014arXiv1409.8262W}, {who observe in their simulations formation of a power-law particle distribution with a super-exponential cut-off}  
\eqb
n'(\gamma)=n'_0\gamma^{-s}e^{-\gamma^2/\gamma'^2_{\rm sl}}, 
\label{distrsl}  
\eqe
where $\gamma'_{\rm sl}$ is the Lorentz factor, for which particle gyroradius becomes equal to a fraction $\zeta$ of {the system size.  
The simulations have not been initialized with a force-free configuration, which is of interest here, and they 
do not include radiative losses, nevertheless we expect that when the escape effects are dominant, a generic type of particle distribution (\ref{distrsl}) is formed, which we adopt here. 
In our case, however, the current sheet contains disordered fields, and locally the acceleration lengthscale is comparable to the particle confinement scale, which we estimate as being roughly the current sheet thickness} $\Delta\lambda = \Delta 2\pi\rl \beta\approx\Delta 2\pi\rl$, thus
\begin{align}
\gamma'_{\rm sl}&= \frac{\gammav\zeta(\Delta2\pi\rl)eB'_{\rm tot}}{mc^2} \\
&\approx 2\times10^{8} \zeta_{-1} \Delta_{-1} \hat{r}_{1}^{-1} \dot{E}_{38}^{1/2} \left(\frac{\Gamma_1^2\hat{r}^{-2}_1+\sin^2\theta}{1+\sin^2\chi}\right)^{1/2}
\label{gammasl}
\end{align}
Here we have taken into account that in the lightcurves 
the width of the pulses normalized to the period of the pulsar, if assumed to reflect the ratio of the sheet thickness to the wind wavelength, implies a parameter $\Delta\approx 0.1$ \citep{2005ApJ...627L..37P}. We keep it the same for each pulsar. However, a physically motivated definition of the current sheet thickness 
is based on the regime in which particle acceleration proceeds. It depends solely on the combination of two pulsar observables: its spin down power and its period, since  $\gamma'_{\rm rad}$ and $\gamma'_{\rm sl}$ become comparable when $\dot{E}_{38}^{3/2}/P_{-2}\sim0.002$ (where according to our fits discussed below $\Gamma_1\approx2.5$, $\hat{r}_1\approx3$).   
For instance, for the Crab pulsar $\dot{E}_{38}^{3/2}/P_{-2}\approx3$ and for the Vela pulsar $\dot{E}_{38}^{3/2}/P_{-2}\approx0.002$. They are typical examples of two different scenarios.  

Synchrotron emission of an accelerated particle can be observed if its gyroradius is larger than the electron inertial length, a scale on which the reconnection takes place \citep{2004PhPl...11.1151J, 2013arXiv1307.7008T}. 
This defines the low energy cut-off $\gamma'_{\rm c}$ in the particle distribution, such that in both eq.~(\ref{distrrad}) and eq.~(\ref{distrsl}) $\gamma>\gamma'_{\rm c}$. In a general form the cut-off is given by $\gamma'_{\rm c}=B'/(8\pi N'mc^2)^{1/2}$, where the particle density is 
\eqb N'=\int_{\gamma'_{\rm c}}^{\infty} n'(\gamma)d\gamma 
\label{totdens}
\eqe
and the parameter $n'_0$ in eq.~(\ref{distrrad}) and eq.~(\ref{distrsl}) is determined by the pressure balance, eq.~(\ref{pbalance}). Calculation of $\gamma'_{\rm c}$ can be simplified by approximating particle distributions, eq.~(\ref{distrrad}) and eq.~(\ref{distrsl}), by a power law with a sharp high energy cut-off at $\gamma'_{\rm rad}$ or $\gamma'_{\rm sl}$, respectively.

Using eq.~(\ref{totdens}) and the relation 
$N'(\bmath{r})=\kappa N_{\rm GJ, LC}\rl^2/\gammav r^2$, 
where $N_{\rm GJ, LC}=\omega\bl/2\pi e c$ is the Goldreich-Julian particle number density at the light cylinder, one can constrain {the number density parameter} $\kappa$ of radiating particles 
$\kappa\approx (e c P \hat{r}/B')\int n'(\gamma)d\gamma$.  
Note that this expression determines only a lower limit on the multiplicity, because the low energy cut-off $\gamma'_{\rm c}$ is an estimation of the lowest energy of radiating particles, whose emission is observed.
\section{Pulsed emission}
\label{sec:Emission}
\subsection{The gamma-ray spectra}
Synchrotron flux produced by a particle distribution $n'(\gamma)$ is given by
\eqb
\epsilon F_{\epsilon}^{\rm syn}=\int_{V}dV\frac{D^3}{4\pi d^2}\frac{\sqrt{3} e^3 B'_{\rm tot}}{h}\epsilon'\int_{\gamma'_{\rm c}}^{\infty} d\gamma' n(\gamma') R(z)
\label{synchrfunction}
\eqe
where $z=(2 \epsilon'B_{\rm cr})/(3B'_{\rm tot}\gamma'^2)$ and $B_{\rm cr}=m^2c^3/e\hbar=4.4\times10^{13}$ G is the quantum critical field, 
\eqb
R(z)=z\int_z^{\infty} d\xi K_{5/3}(\xi)\approx1.78 z^{0.297} e^{-z}.
\eqe
$\epsilon$ and $\epsilon'$ is the photon energy normalized to the electron rest mass in the lab and in the comoving frame, respectively; $d$ is the distance between the observer and the pulsar, and $D$ is the Doppler factor of a wind (for details see \citet{2005ApJ...627L..37P}). 
Introducing $y=\gamma'/\gamma'_{\rm rad}$ and using the particle spectrum (\ref{distrrad}), 
we obtain
\eqb 
\epsilon F_{\epsilon}^{\rm syn} \propto x^{1.3}\int_{\gamma'_{\rm c}/\gamma'_{\rm rad}}^{\infty} dy e^{-(s+0.6)\ln y-y-x/y^2}, 
\eqe
where $x=(2 \epsilon'B_{\rm cr})/(3B'_{\rm tot}\gamma'^2_{\rm rad})$. 
Asymptotic behaviour of this integral can be calculated by the steepest descent method. 
We find an extremum $y_0$ of the integrand from the condition of the vanishing of its first derivative. In the exponential tail $\gamma'\gg\gamma'_{\rm rad}$, the solution can be found analytically $y_0\approx(2x)^{1/3}$. Next, we expand the integrand to the second order around $y_0$. Assuming $\gamma'_{\rm c}\ll\gamma'_{\rm rad}$, we integrate a resulting Gaussian function. An asymptotics takes the form 
\eqb
\epsilon F_{\epsilon}^{\rm syn} \propto x^{1.3-(s+0.6)/3}e^{-1.9\,x^{1/3}} \label{asympexp}
\eqe
Note that this is very similar to the spectrum $\propto \exp(x^{0.35})$ obtained by \citet{2013A&A...550A.101A} by a fit.

In a similar way we calculate an asymptote of a spectrum produced by the particle distribution (\ref{distrsl}). In this case the exponent has a form $f(y)=-(s+0.6)\ln y-y^2-x/y^2$ and its derivative has a maximum at $y_0\approx x^{1/4}$. 
The flux has the following asymptotic behaviour:
\eqb
\epsilon F_{\epsilon}^{\rm syn}\propto x^{1.3-(s+0.6)/4}e^{-2\,x^{1/2}} \label{asympsuperexp}
\eqe 
The asymptotes (\ref{asympexp}) and (\ref{asympsuperexp}), together with fitted spectra (see next section) are shown in Fig.~\ref{fig1} and Fig.~\ref{fig2}, respectively.
\subsection{Constraining the model parameters}
\label{sec:fitting}

\begin{figure}
\scalebox{1}{
\input{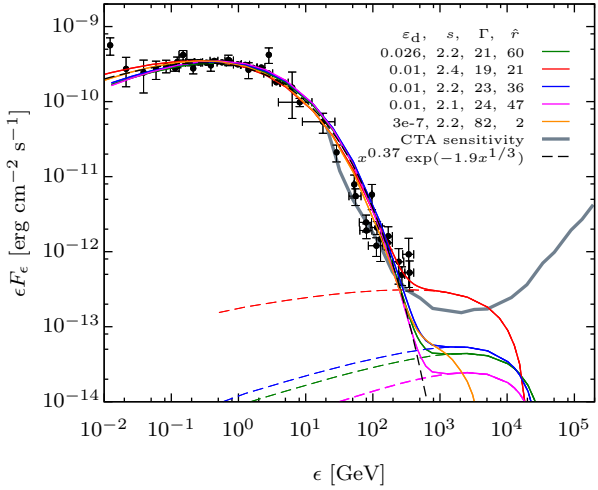}}
\caption{\small Several best fits to the Crab spectrum. Thick gray line shows the predicted CTA sensitivity \citep{emma}, black dashed line is the asymptote (\ref{asympexp}), 
plotted with $s=2.2$.
Dashed lines show SSC components, while solid lines are the total (synchrotron + SSC) spectrum. Black points are the data from \citet{2001A&A...378..918K}, \citet{2010ApJ...708.1254A} and \citet{2014A&A...565L..12A}.}
\label{fig1}
\end{figure}

\begin{figure}
\scalebox{1}{
\input{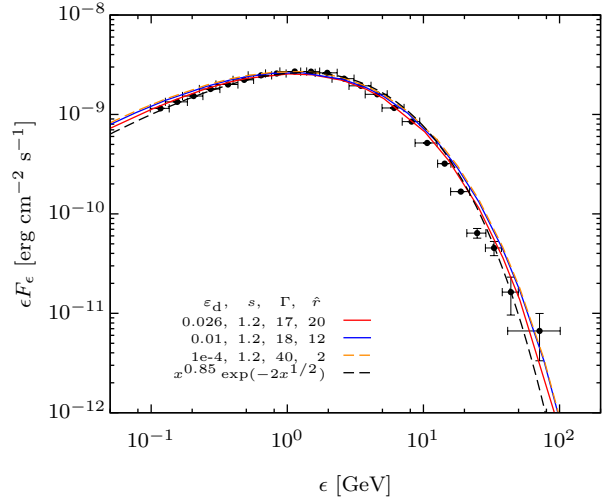}}
\caption{\small Several best fits to the synchrotron spectrum of Vela. For a hard particle index, the SSC component is weaker by several orders of magnitude and it is not shown in the plot. Black dashed line is the asymptote (\ref{asympsuperexp}), plotted with $s=1.2$. 
Black points are the data from Fermi-LAT \citep{2014arXiv1410.5208L}}
\label{fig2}
\end{figure}
 
There are four free parameters of the model: the Lorentz factor of the wind $\gammav$, the dissipation distance in units of the light cylinder $\hat{r}$, the particle index $s$ and the dissipation efficiency $\varepsilon_{\rm d}$. 
We constrain them by matching the synchrotron flux (\ref{synchrfunction}) to the phase-averaged spectra. 

To compute the fluxes in our model we integrate the wind emissivity over 3D volume (see eq.~(\ref{synchrfunction})) using spectral methods \citep{1986nras.book.....P}. According to this approach,  
the integrand is expanded in a basis of orthogonal polynomials  
and integrated term by term.  
As bases we choose the same special functions as proposed by \citet{2013MNRAS.434.2636P}, i.e., Chebyshev polynomials in coordinate $r$, Chebyshev polynomials in $\cos\theta$, and a Fourier series in $\phi$. 
This choices allow us to use a fast cosine transform to calculate the coefficients of expansion in the polynomial bases, and to apply directly the Clenshaw-Curtis quadrature rules for computing integrals (Press et al. 1986). 

The model spectrum is calculated 
at the peak of the lightcurve, and we obtain the phase averaged flux
using $F_{\rm av} = F_0 I$, where $F_0$ is the flux at the peak maximum,
and $I$ is the integral area under the lightcurve (assuming  
that the emission peaks are normalized to 1). This approach is justified by the fact that the model spectra change very weakly with the pulsar phase (due to a weak angle dependence of the Doppler factor when the emission is boosted to the observer frame). 

In Fig.~\ref{fig1} we present several fits to the Crab data from Fermi-LAT and MAGIC. 
{There is no unique solution as long as the measurements in the TeV range are missing. Computed spectra result from an interplay between all four parameters of the model, which must be found simultaneously in order to give a reasonable fit. 
The green, blue and orange curves show the spectra with $s=2.2$ and different $\varepsilon_{\rm d}$, for which we have found the best-fit $\Gamma$ and $\hat{r}$. 
In general, smaller $\varepsilon_{\rm d}$ lead to smaller $\hat{r}$ and larger $\Gamma$, thus, by increasing $\Gamma$ and decreasing $\epsilon_{\rm d}$ we are able to fit the data with the model of emission close to the pulsar, at $\hat{r}=2$ (orange curve). This fit gives an estimate of the maximum possible Lorentz factor of the wind, on the order of $\Gamma\sim100$, similar to the value assumed in previous studies} \citep{2012ApJ...745..108A,2013ApJ...771...53M}. {Note that for the best fits with the same value of $s$ the brightness of the SSC component does not change significantly, but the maximum energy of SSC photons does. It is estimated by the Klein-Nishina limit, boosted to the observer frame:} \eqb \epsilon\approx 2\Gamma \gamma'_{\rm rad}mc^2\approx3.6\, \Gamma_{1}^{5/2}P_{-2}^{1/2}\varepsilon_{\rm d,-2}^{1/2}\dot{E}_{38}^{-1/4} {\rm TeV}. \eqe 

{Red, blue and pink curves show the fits for $\varepsilon_{\rm d}=0.01$ and different $s$, for which, as before, we have also found the best-fit $\Gamma$ and $\hat{r}$. In this case, the larger is $s$, the smaller are also $\hat{r}$ and $\Gamma$. The brightness of the SSC component changes significantly, being greater for larger $s$. With the currently available VHE data the upper limit on the particle index can be estimated to be $s\approx2.4$ (red curve).} 
In general, soft indices $s\ga2$ are required to reproduce the gamma-ray bump in the Crab spectrum.  
The low energy cut-off in the particle distribution is estimated to be roughly $\gamma'_{\rm c}\sim50$. 

A fit to the Vela spectrum requires a much harder index $s\approx1.2$ (Fig.~\ref{fig2}). In this case the SSC component is weaker by several orders of magnitude (not shown in the plot), because the energy density resides mainly in the highest energy particles, which do not produce enough of low energy synchrotron photons that can be upscattered to very high energies (VHE). 
The low energy cut-off in the particle distribution is roughly $\gamma'_{\rm c}\approx2\times10^3$. 

{The number density $\kappa$ of the radiating particles and the reconnection rate $\tau$ for both pulsars are compared in Table~\ref{tableone} for two (exemplary) best-fit sets of parameters: for the efficiency $\varepsilon_{\rm d}=0.01$ and for the dissipation distance $\hat{r}=2$. In all cases our fits suggest that the number density of radiating particles is much smaller in the Vela wind than in the Crab wind, but the reconnection rate is higher. According to simulations} \citep{2013ApJ...775...50T,  2014PhRvL.113o5005G} this result suggests that the Vela wind is more magnetized.   
\begin{table}
\begin{tabular}{c|c|c || c|c}
& Crab & Vela &  Crab & Vela \\ \hline  
$\varepsilon_{\rm d}$ & {\bf 0.01} & {\bf 0.01} & $\varepsilon_{\rm d}=3\times10^{-7}$ & $\varepsilon_{\rm d}=10^{-5}$\\
$\Gamma$ & 23 & 18 & 82 & 40 \\
$\hat{r}$ & 36 & 12 & {\bf 2} & {\bf 2} \\
$\tau$ & 0.18 & 0.26 & 0.001 & 0.01 \\
$\kappa$ & $7\times10^4$ & 14 & 0.5 & 0.05
\end{tabular}
\caption{Reconnection rate $\tau$ and the number density $\kappa$ of radiating particles obtained for two exemplary fits to the spectra.}
\label{tableone}
\end{table}

\section{Conclusions}
\label{sec:Conclusion}
The spectra of gamma-ray pulsars give insight into the physics of relativistic reconnection in pulsar winds. 
Particle acceleration at the reconnection sites can proceed either in the radiative cooling regime (in the most powerful pulsars with $\dot{E}_{38}^{3/2}/P_{-2}\ga 0.002$), or in the size limited regime if the pulsar is less energetic. 
The synchrotron spectra behave asymptotically like $\propto\exp(\epsilon^{1/3})$ 
and $\propto\exp(\epsilon^{1/2})$, respectively. 

In the Crab spectrum, a new SSC component at tens of TeV is expected to be  
{observed by} the upcoming experiment CTA \citep{emma}. 
The MAGIC Collaboration \citep{crabtev} has recently obtained new measurements of the Crab pulsar at $\sim 2$ TeV. A comparison of our results with the spectrum in the VHE range will give a constraint on the particle index $s$ of the population accelerated during reconnection in the wind. 
  
We put {an upper limit on the Lorentz factor of the Crab wind $\Gamma\la100$ and of the Vela wind $\Gamma\la50$, which are obtained if the emission occurs at $r_{\rm diss}=2\rl$.} These values are in good agreement with previous works on optical polarization signatures \citep{2005ApJ...627L..37P} and Crab flares \citep{2013MNRAS.436L..20B}. The {number density of radiating particles in the wind depends on the dissipation distance, but in each case our fits to the spectra imply that it is smaller in the Vela wind than in the Crab wind. The estimated reconnection rate is larger for Vela.} 

Our study implies that the current sheet structure is imprinted on the pulsar gamma-ray lightcurves. Future modelling of the pulse profile change with energy, together with the phase-resolved spectroscopy, will give much more insight into the reconnection dynamics in pulsar winds. 
\section*{Acknowledgments}
We are grateful to John Kirk and to the anonymous referee for helpful comments on the manuscript. We also thank Alice Harding for drawing our attention to the newest results of MAGIC. This work has been supported by the French National Research Agency (ANR) through the grant No. ANR-13-JS05-0003-01 (project EMPERE). We also benefited from the computational facilities available at Equip@Meso (Universit\'e de Strasbourg).
\bibliographystyle{mn2e}

\begin{thebibliography}{36}
\providecommand{\natexlab}[1]{#1}

\bibitem[{{Abdo} et~al.(2010)}]{2010ApJ...708.1254A}
{Abdo}, A.~A., et~al., 2010, \apj, 708, 1254

\bibitem[{{Ackermann} et~al.(2013)}]{2013ApJS..209...34A}
{Ackermann}, M., et~al., 2013, \apjs, 209, 34

\bibitem[{{Aharonian}, {Bogovalov} \&
  {Khangulyan}(2012)}]{2012Natur.482..507A}
{Aharonian}, F.~A., {Bogovalov}, S.~V., \& {Khangulyan}, D., 2012, \nat, 482, 507

\bibitem[{{Aleksi{\'c}} et~al.(2012)}]{2012A&A...540A..69A}
{Aleksi{\'c}}, J., et~al., 2012, \aap, 540, A69

\bibitem[{{Aleksi{\'c}} et~al.(2014)}]{2014A&A...565L..12A}
{Aleksi{\'c}}, J., et~al., 2014, \aap, 565, L12

\bibitem[{Aliu et~al.}(2011)]{2011Sci...334...69V}
{Aliu}, E., et~al. (VERITAS Collaboration), 2011, Science, 334, 69

\bibitem[{{Arka \& Dubus}(2013)}]{2013A&A...550A.101A}
{Arka}, I., \& {Dubus}, G., 2013, \aap, 550, A101

\bibitem[{{Arka \& Kirk}(2012)}]{2012ApJ...745..108A}
{Arka}, I., \& {Kirk}, J.~G., 2012, \apj, 745, 108

\bibitem[{{Bai \& Spitkovsky}(2010)}]{2010ApJ...715.1282B}
{Bai}, X.~N., \& {Spitkovsky}, A., 2010, \apj, 715, 1282

\bibitem[{{Baty, P\'etri \& Zenitani}(2013){Baty}, {P\'etri} \&
  {Zenitani}}]{2013MNRAS.436L..20B}
{Baty}, H., {P\'etri}, J., \& {Zenitani}, S., 2013, \mnras, 436, L20

\bibitem[{{Coroniti}(1990)}]{1990ApJ...349..538C}
{Coroniti}, F.~V., 1990, \apj, 349, 538

\bibitem[{{Giannios}(2013)}]{2013MNRAS.431..355G}
{Giannios}, D., 2013, \mnras, 431, 355

\bibitem[{{Guo} et~al.(2014){Guo}, {Li}, {Daughton} \&
  {Liu}}]{2014PhRvL.113o5005G}
{Guo}, F., {Li}, H., {Daughton}, W., {Liu}, \& Y.~H., 2014, Phys. Rev. Lett.,
  113, 155005

\bibitem[{{Harrison} \& {Neukirch}(2009)}]{2009PhRvL.102m5003H}
{Harrison}, M.~G., \& {Neukirch}, T., 2009, Phys. Rev. Lett., 102, 135003

\bibitem[{{Jaroschek} et~al.(2004){Jaroschek}, {Treumann}, {Lesch} \&
  {Scholer}}]{2004PhPl...11.1151J}
{Jaroschek}, C.~H., {Treumann}, R.~A., {Lesch}, H., \& {Scholer}, M., 2004, Phys. of
  Plasmas, 11, 1151

\bibitem[{Kalapotharakos}, {Harding} \& {Kazanas}(2014)]{2014ApJ...793...97K}
{Kalapotharakos}, C., {Harding}, A.~K., \& {Kazanas}, D., 2014, \apj, 793, 97

\bibitem[{{Kirk}, {Skj{\ae}raasen} \& {Gallant}(2002){Kirk}, {Skj{\ae}raasen} \&
  {Gallant}}]{2002A&A...388L..29K}
{Kirk}, J.~G., {Skj{\ae}raasen}, O., \& {Gallant}, Y.~A., 2002, \aap, 388, L29

\bibitem[{{Kuiper} et~al.(2001)}]{2001A&A...378..918K}
{Kuiper}, L., {Hermsen}, W., {Cusumano}, G., {Diehl}, R., {Sch{\"o}nfelder}, V.,
  {Strong}, A., {Bennett}, K., \& {McConnell}, M.~L., 2001, \aap, 378, 918

\bibitem[{{Leung} et~al.(2014){Leung}, {Takata}, {Ng}, {Kong}, {Tam}, {Hui} \&
  {Cheng}}]{2014arXiv1410.5208L}
{Leung}, G.~C.~K., {Takata}, J., {Ng}, C.~W., {Kong}, A.~K.~H., {Tam}, P.~H.~T.,
  {Hui}, C.~Y., \& {Cheng}, K.~S., 2014, \apjl, 797, L13

\bibitem[{{Lyubarskii}(1996)}]{1996A&A...311..172L}
{Lyubarskii} Y.~E., 1996, \aap, 311, 172

\bibitem[{{Lyubarsky} \& {Kirk}(2001)}]{2001ApJ...547..437L}
{Lyubarsky}, Y., \& {Kirk}, J.~G., 2001, \apj, 547, 437

\bibitem[{{Lyutikov}(2013)}]{2013MNRAS.431.2580L}
{Lyutikov}, M., 2013, \mnras, 431, 2580

\bibitem[{{Mochol} \& {Kirk}(2013)}]{2013ApJ...771...53M}
{Mochol}, I., \& {Kirk}, J.~G., 2013, \apj, 771, 53

\bibitem[{de O{\~n}a-Wilhelmi et al.}(2012)]{emma}
{de O{\~n}a-Wilhelmi}, E., et al, 2012, Astropart. Phys., 43, 287

\bibitem[{{P{\'e}tri}(2012)}]{2012MNRAS.424.2023P}
{P{\'e}tri}, J., 2012, \mnras, 424, 2023

\bibitem[{{P{\'e}tri}(2013)}]{2013MNRAS.434.2636P}
{P{\'e}tri}, J., 2013, \mnras, 434, 2636

\bibitem[{{P{\'e}tri} \& {Kirk}(2005)}]{2005ApJ...627L..37P}
{P{\'e}tri}, J., \& {Kirk}, J.~G., 2005, \apjl, 627, L37

\bibitem[{{Press}, {Flannery} \& {Teukolsky}(1986)}]{1986nras.book.....P}
{Press}, W.~H., {Flannery}, B.~P., \& {Teukolsky}, S.~A., Numerical recipes. The art of scientific computing., Cambridge: University Press, 1986

\bibitem[{{Sironi} \& {Spitkovsky}(2014)}]{2014ApJ...783L..21S}
{Sironi}, L., \& {Spitkovsky}, A., 2014, \apjl, 783, L21

\bibitem[{{Spitkovsky}(2006)}]{2006ApJ...648L..51S}
{Spitkovsky}, A., 2006, \apjl, 648, L51

\bibitem[{{Takamoto}(2013)}]{2013ApJ...775...50T}
{Takamoto}, M., 2013, \apj, 775, 50

\bibitem[{{Treumann} \& {Baumjohann}(2013)}]{2013arXiv1307.7008T}
{Treumann} R.~A., \& {Baumjohann} W., 2013, arXiv: 1307.7008

\bibitem[{{Werner} et~al.(2014){Werner}, {Uzdensky}, {Cerutti}, {Nalewajko} \&
  {Begelman}}]{2014arXiv1409.8262W}
{Werner}, G.~R., {Uzdensky}, D.~A., {Cerutti}, B., {Nalewajko}, K., \& {Begelman},
  M.~C., 2014, arXiv:1409.8262

\bibitem[{{Zanin et al.}(2014)}]{crabtev}
{Zanin, R., for the MAGIC Collaboration}, 2014, 5th Int. Fermi Symp.

\end{thebibliography}

\label{lastpage}
\end{document}